\documentclass[Afour,sageh,times]{sagej}
\usepackage{natbib}
\setcitestyle{numbers,square}
\usepackage{comment}
\usepackage{moreverb,url}

\usepackage[colorlinks,bookmarksopen,bookmarksnumbered,citecolor=red,urlcolor=red]{hyperref}

\newcommand\BibTeX{{\rmfamily B\kern-.05em \textsc{i\kern-.025em b}\kern-.08em
T\kern-.1667em\lower.7ex\hbox{E}\kern-.125emX}}

\def\volumeyear{2026}

\usepackage{listings}
\usepackage{color}
\definecolor{dkgreen}{rgb}{0,0.6,0}
\definecolor{gray}{rgb}{0.5,0.5,0.5}
\definecolor{mauve}{rgb}{0.58,0,0.82}

\newcommand\todo[1]{\textcolor{red}{#1}}
\newcommand\toauthor[1]{\textcolor{blue}{#1}}
\newcommand\response[1]{\textcolor{green}{#1}}

\usepackage{algorithmic}
\usepackage{subfig}
\usepackage{graphicx}
\usepackage{textcomp}
\usepackage{xcolor}
\usepackage[labelfont=bf,textfont=bf]{caption}
\usepackage[strings]{underscore}

\graphicspath{{images/}}

\begin{document}

\runninghead{Joshi and Vadhiyar}

\title{FTHP-MPI: Towards Providing Replication-based Fault Tolerance in a Fault-Intolerant Native MPI Library}

\author{Sarthak Joshi\affilnum{1} and Sathish Vadhiyar\affilnum{1}}

\affiliation{\affilnum{1}Indian Institute of Science, Bengaluru}

\corrauth{Sarthak Joshi, Department of Computational and Data Sciences
Indian Institute of Science,
Bengaluru,
Karnataka, India}

\email{sarthakjoshi@iisc.ac.in}


\begin{abstract}
Faults in high-performance systems are expected to be very frequent in the current exascale computing era. To compensate for a higher failure rate, the standard checkpoint/restart technique would need to create checkpoints at a much higher frequency, resulting in an excessive amount of overhead, which would not be sustainable for many scientific applications. To improve application efficiency in such high-failure environments, the mechanism of replication of MPI processes was proposed. Replication allows for fast recovery from failures by simply dropping the failed processes and using their replicas to continue the regular operation of the application.

In this paper, we have implemented {\it FTHP-MPI} (Fault Tolerance and High Performance MPI), a novel fault-tolerant MPI library that augments checkpoint/restart with replication to provide resilience from failures. The novelty of our work is that it is designed to provide fault tolerance in a native MPI library that does not provide support for fault tolerance.
This lets application developers achieve fault tolerance at high failure rates while also using efficient communication protocols in the native MPI libraries that are generally fine-tuned for specific HPC platforms. We have also implemented efficient parallel communication techniques that involve replicas. Our framework deals with the unique challenges of integrating support for
checkpointing and partial replication.

We conducted experiments with three applications, HPCG, PIC, and CloverLeaf. We show that, for large-scale systems where failure intervals are expected to be within an hour, our replication-based library achieves higher efficiency and performance than checkpoint-based approaches. We show that, under failure-free conditions, the additional overheads from replication are negligible in our library.

\end{abstract}


\maketitle

\section{Introduction}

Large-scale systems are prone to failures due to both hardware and software faults. The standard method to handle failures is the checkpoint/restart mechanism\cite{hargrove-berkeley-2006}\cite{sankaran-lam/mpi-2005}. In this method, the application state is saved as checkpoints at regular intervals. Upon failure, the application is restarted, the last saved checkpoint is used to recover the saved state, and execution continues from that point. However, as Exascale systems are being built, the failure rate is expected to considerably increase due to the complexities of the components and the interconnections. Checkpoints need to be created at a higher frequency to compensate for the high failure rates. Furthermore, the saved checkpoints will be loaded much more frequently as a restart will be needed at every failure\cite{benoit-replication-2019}. This results in large overheads that will result in significant performance loss for many scientific applications\cite{walters-replication-based-2009}.

Hence, fault tolerance using replication was proposed\cite{ferreira-evaluating-2011}\cite{benoit-replication-2019}\cite{bougeret-using-2014}\cite{george-fault-2015}. In this strategy, replica processes are maintained for the original set of processes. These replica processes maintain the same application state as the original processes. Replication can provide faster recovery from failures by simply dropping the failed processes and continuing with the replicas. This helps increase the mean time to interruption (MTTI) of the application since both the original and its corresponding replica process have to fail for the application to fail. This also allows using longer checkpoint intervals.
Most of the existing fault-tolerant MPI libraries do not harness the efficient native MPI communications that are highly tuned to the network topology and other hardware aspects, thereby compromising performance.

In this paper, we have developed {\it FTHP-MPI}, an MPI library based on augmenting checkpoint/restart with replication. Our library allows various degrees of replication to be used for partial replication. The novelty of our library is that it provides both fault tolerance, by utilizing coordinated checkpointing along with replication, and high performance, by utilizing a native MPI library for communications, thereby providing the best of both worlds. We have created an interface with which a native MPI library (both open and closed source) can be loaded and used for communications. Our framework effectively hides all process failure incidents from the native MPI library.

We have also implemented parallel and efficient communication strategies involving both the computational and replica processes while utilizing the communications of the native MPI library. We implement portable mechanisms for carefully mapping address spaces and resetting pointers to create and manage live replication. Our FTHP-MPI framework also deals with the unique challenges of integrating support for checkpointing and partial replication, including the challenge of different numbers of processes used for checkpointing and restart during application failure. We also efficiently repair communicators during a failure using a complete non-blocking approach without requiring all the processes to enter an MPI function and receive an error code, unlike ULFM.

We conducted experiments with three applications, HPCG, PIC, and CloverLeaf. We show that for large-scale systems where the failure intervals are expected to be within an hour, our replication-based library provides higher efficiency and performance than checkpointing-based approaches. We show that beyond a certain number of processors, better efficiency is obtained by using the additional processors for replication than using all the processors for application execution with checkpointing. This, we show, is because replication mostly does useful work even though half of it is redundant. We show that under failure-free conditions, the additional overheads, mainly due to communications needed for maintaining the replicas, are negligible in our library.

The following are the primary contributions of our paper.
\begin{enumerate}
    \item We propose the first framework that provides replication-based fault tolerance for a native MPI library that does not have support for fault tolerance, thereby combining fault tolerance with high performance provided by the native MPI libraries. We have designed our library in a way that allows the user to utilize the efficient implementations of the native MPI libraries while also having replication-based fault tolerance without any changes to the code bases of the native MPI libraries.
    \item We implement a unified fault tolerance framework that combines checkpoint/restart with replication and addresses various challenges for these to work together.
    \item We performed experiments with HPCG, PIC, and CloverLeaf applications. Ours is the first work that shows the benefits of using only replication over checkpointing for fault tolerance using actual executions.
\end{enumerate}

Section \ref{related-work} covers related work on techniques for fault tolerance, including checkpointing, algorithm-based fault tolerance, and replication. Section \ref{framework} presents our unified framework for fault tolerance that integrates both the checkpointing and replication techniques. Section \ref{using-multi} describes our methods for providing fault tolerance to a native MPI library that does not support fault tolerance. Section \ref{library} details the implementation aspects of our FTHP-MPI library. Section \ref{failure-mgmt} gives details on the failure management mechanisms of our library. Section \ref{exp-res} gives experiments and results on efficiency, performance, and overheads, comparing replication and checkpointing. Section \ref{con-fut} gives conclusions and future work.

\section{Related Work}
\label{related-work}

Over the years, various approaches have emerged to counter the issue of frequent failures in large-scale systems. Fail-stop failures, in which a processor crashes, are the primary concern of our work. However, there are also other kinds of failures, such as network failures, which need to be dealt with on a link level, and silent errors that corrupt the memory of the application without an immediate crash, resulting in either incorrect outputs or an eventual crash much later in the execution. We now review the related work on providing fault tolerance in large-scale systems.

\subsection{Checkpoint–Restart}
Checkpoint-Restart\cite{hargrove-berkeley-2006}\cite{sankaran-lam/mpi-2005}\cite{andrijauskas-criu-2024}\cite{egwutuoha-survey-2013} is currently the most widely accepted method to handle failures in MPI. There have been various approaches proposed to improve the efficiency of checkpointing techniques to deal with crash failures through the use of multilevel checkpointing\cite{moody-design-2010} that enables the use of multiple types of checkpoints in a single run with varying costs. For example, checkpoints could be written to the RAM, a node-local storage, or to a parallel file system with varying levels of I/O time and resilience.
Other solutions in this category involve checkpointing without global synchronization\cite{guermouche-uncoordinated-2011}, providing failure containment by dividing the application processes across independent clusters\cite{guermouche-hydee:-2012}, and techniques for efficient checkpoint recovery\cite{riesen-alleviating-2012}. There have also been some advancements in failure prediction techniques that can allow more proactive checkpointing\cite{bouguerra-improving-2013}. Tools like DMTCP\cite{ansel-dmtcp-2007} allow checkpoints to be created in an MPI-agnostic manner. MANA\cite{garg-mana-2019} further extends DMTCP to support network-agnostic checkpointing through a split-process approach. This framework runs two separate address spaces inside a single process and switches between them during MPI function call boundaries. Only the address space used outside the MPI function call is checkpointed. Therefore, the MPI library and its network components can be freely slotted in and out as modules independent from the checkpointed address space. 

Our checkpoint/restart mechanism reads and writes memory segments to a checkpoint file, providing transparent user-level checkpointing similar to DMTCP and MANA. DMTCP is primarily useful for checkpointing serial applications with limited MPI support. MANA is a DMTCP extension intended for scalable MPI applications. Similar to these implementations, we use a coordinated checkpointing approach in which a coordinator process is launched to initiate periodic checkpoints across all processes. However, we use a custom coordinator implementation for scalability and to enable failure detection and propagation. Since our work supports the use of replicas alongside checkpointing, not every failure necessarily results in a restart. Therefore, failure information must be communicated to the remaining processes, unlike in other checkpointing implementations, which simply restart upon each failure. As a result, our server-client implementation between the coordinator and the MPI processes involves more than just signal passing and handling, which is not easily achievable with existing coordinator implementations in these pure checkpointing frameworks. Our implementation also addresses unique challenges posed by combining the checkpoint/restart and replication frameworks, such as differences in the number of processes that can participate in checkpointing and restarting, while ensuring that processes remain unique within MPI communications. Addressing this requires a mechanism to partially but transparently checkpoint only the application space data, which is not feasible in an implementation like DMTCP, which relies on full memory dumps. While the split-process approach of MANA makes such a mechanism technically possible, maintaining two full sets of memory mappings in each process can result in substantial memory overhead at scale, especially as our framework also relies on sender-based message logging to handle failures. Therefore, we have chosen to use a custom checkpointing implementation that seamlessly integrates with our replication mechanism.

\subsection{Global Restart}
The Global restart model was proposed to reduce the overhead from checkpoint recovery in bulk synchronous applications\cite{laguna-evaluating-2014}. In this model, instead of aborting the job on failure and restarting from the last saved checkpoint, the application state is restored in the live processes through a rollback mechanism. This saves the costs of reallocating the resources and redeploying the entire application on all the nodes. Reinit\cite{sadayappan-reinit++-2020} is an implementation of this model, which assumes a non-shrinking behavior in which only the failed processes are respawned, and their application states are quickly restored with a coordinated effort by the live processes. Fenix\cite{gamble-specification-2016} is another implementation that also supports the shrinking recovery, where the failed processes are simply dropped, and execution continues with only the remaining number of processes. MPI Stages\cite{sultana-mpi-2018}\cite{schafer-extending-2020} further introduces an MPI state checkpoint that can be used to bypass the MPI state creation, involving communicator creation, custom datatype creation, etc., for an even faster recovery in the respawned processes. Our replication-based fault tolerance is a form of the global restart model where the application is not aborted on failure and instead attempts to continue with the help of the replicas. Unlike the earlier global restart works, our work does not require the application program to be modified.

\subsection{Error-code based recovery}
The error-code-dependent recovery model returns control to the application upon failure instead of aborting and allows it to handle the failures suitably. User Level Failure Mitigation (ULFM)\cite{bland-post-failure-2013} is based on such a model and was proposed by the Fault Tolerant Working Group of the MPI Forum. Here, error codes are returned by the MPI routines to notify the application of failures. The application can then subsequently utilize other MPI library-provided routines to transfer control to an error handler, prune the communicators of failed processes, and perform any other recovery mechanisms that it needs to continue execution. One major issue with ULFM's adoption is the need to rebalance the load across the remaining processes\cite{laguna-evaluating-2014}. Local Failure Local Recovery (LFLR)\cite{teranishi-toward-2014} is a ULFM-based model that uses spare processes that are activated upon failure to keep the number of processes constant. However, the spare processes essentially operate on a skeleton code in this implementation without holding any actual data, which can require heavy manipulation on the application side. Furthermore, a set of live processes needs to redundantly hold the data in their user-space memory so that they can coordinate to recover the state of the activating process. Depending on the level of this data redundancy, if multiple processes from the same group die, the spare processes can no longer be activated.

The ULFM framework only mandates that the application does not automatically abort upon failure and that MPI functions return error codes to the user. Failure handling is left entirely to the user and often requires application- or algorithm-level support to truly provide fault tolerance, as with LFLR. Our work uses the communicator-shrinking concept from ULFM to continue execution as long as at least one replica of each process remains alive. Replica processes execute the same code rather than a skeleton. This eliminates the need for recovery mechanisms besides lost communications, as all the data is already present in the replicas. Unlike the standard ULFM protocol, our implementation does not rely on returning error codes. By performing communicator shrinking in the background immediately after the failure is propagated, our implementation handles failures without user intervention and without requiring any changes to the application code.

\subsection{Replication}
Replication involves creating copies or replicas of a process that redundantly performs the same operations. Upon failure, as long as one copy of each process survives, the job can continue its execution. Replication-based solutions have seen a growing interest as some studies have shown the impact of replication on the mean time to application interruption\cite{benoit-replication-2019}\cite{bougeret-using-2014}\cite{casanova-impact-2015} using simulations. As we scale up to a high number of processes, the mean time for at least one process to fail decreases rapidly, but the average time for exactly two copies of a process to fail decreases at a much slower rate. The application can continue execution through multiple failures as long as one replica survives. Therefore, using replicas allows us to use higher checkpoint intervals and, thus, reduce the checkpoint overhead. There exists a critical point where the high number of processes drives the failure rate high enough that using replication becomes preferable, despite it being only 50\% as efficient due to redundancy.

Much of the attention in this field is either on utilizing replicas to identify and recover from inconsistencies due to soft errors, like with RedMPI\cite{fiala-detection-2012}, or on getting past the 50\% efficiency threshold. Increasing application efficiency in the presence of replication has been attempted by utilizing partial replication\cite{stearley-does-2012} that only replicates some of the processes. Efficiency can also be improved by sharing the work between two replicas to reduce redundancy\cite{ropars-efficient-2015}\cite{samfass-teampireplication-based-2020}. By exploiting application properties, replicas can divide the work among themselves and then ensure that the output of the work is shared at critical points in the application. Furthermore, it has been shown that accurate failure predictions, along with adaptive replication, can greatly improve the efficiency\cite{george-fault-2015}. By identifying which processes are at a high risk of failure and adaptively switching replicas across them, it is possible to tolerate a large number of failures using a small number of replicas. While there have been some simple MPI libraries like rMPI \cite{ferreira-evaluating-2011} that have implemented replication, these implementations are generally deeply tied to a single MPI library and involve heavy modifications in its source code to work. While this work demonstrates low replication overheads under failure-free conditions with a real implementation, it only shows simulated comparisons to the checkpoint/restart mechanism under failure conditions.
To our knowledge, ours is the first work that provides fault tolerance using both checkpoint/restart and replication while harnessing efficient communications in the native MPI libraries and providing this in an MPI-agnostic manner that is independent of the application code. Instead of being fully or partially simulation-based, we have designed a complete end-to-end pipeline for handling failures in real applications, involving failure detection, propagation, and handling, which is all seamlessly integrated into the MPI standard and thoroughly tested to account for all the edge cases in failure scenarios.
\section{A Unified Fault Tolerance Framework for Checkpoint/Restart and Replication}
\label{framework}

By utilizing both checkpointing and replication together, we can reduce the time overhead in checkpoint/restart significantly. The probability of both a process and its replica failing is lower than a single-process failure, resulting in lower application failure rates with replication than with checkpointing alone. Therefore, using replicas allows us to use a higher checkpoint interval that reduces checkpoint overhead. Note that checkpointing will still be needed, especially in environments with high application failure rates, with large probabilities of failures of both a process and its replica.

However, integrating checkpointing and replication in a unified MPI-agnostic fault tolerance framework is challenging. For example, we need replica processes to be equivalent to their original counterparts but still be unique, such that the relevant MPI communications can be performed as if they were separate processes. This is especially important as we are performing communications using a native MPI library. If we implement replication in a similar manner as standard checkpointing and just copy over all memory segments, it will lead to the native MPI memory segment also getting copied over, which would result in the native MPI library being unable to identify all the processes uniquely, as half of them would have the same internal data, including any identifiers.
There could also be a difference in the number of processes creating a checkpoint and those restoring from the checkpoint. This section describes our methods for addressing these challenges.

\subsection{Checkpoint/Restart Mechanism}

The creation of a checkpoint/restart framework involves multiple stages. The first stage is to set up an MPI environment with checkpointing enabled. We use a system of {\it checkpointing coordinators} for this purpose. The second stage is about ensuring all the processes reach a safe state before checkpoint writing or reading. The next stage is the process of safely dumping or reading the relevant data to or from the checkpoint file.

\subsubsection{Checkpointing Coordinators}

Our library uses a coordinated checkpointing approach that uses a set of checkpoint coordinators to synchronously trigger a checkpoint write operation in all the processes. We use similar mechanisms as DMTCP\cite{ansel-dmtcp-2007} to read and write memory mappings. We have augmented these mechanisms such that they also support safely restoring in cases where some of the mappings overlap with the code executing the restore operation. For executing across multiple nodes, we launch a checkpoint coordinator in each node that communicates with the processes local to that node and with the checkpoint coordinators of the other nodes. For periodic checkpointing, the checkpoint timer is only run on a single primary coordinator that messages the other coordinators when the timer completes, following which signals are sent to the local processes.

\subsubsection{Setting up an MPI Environment with Checkpointing Coordinators}

\begin{figure}
    \centering
    \includegraphics[width=\linewidth]{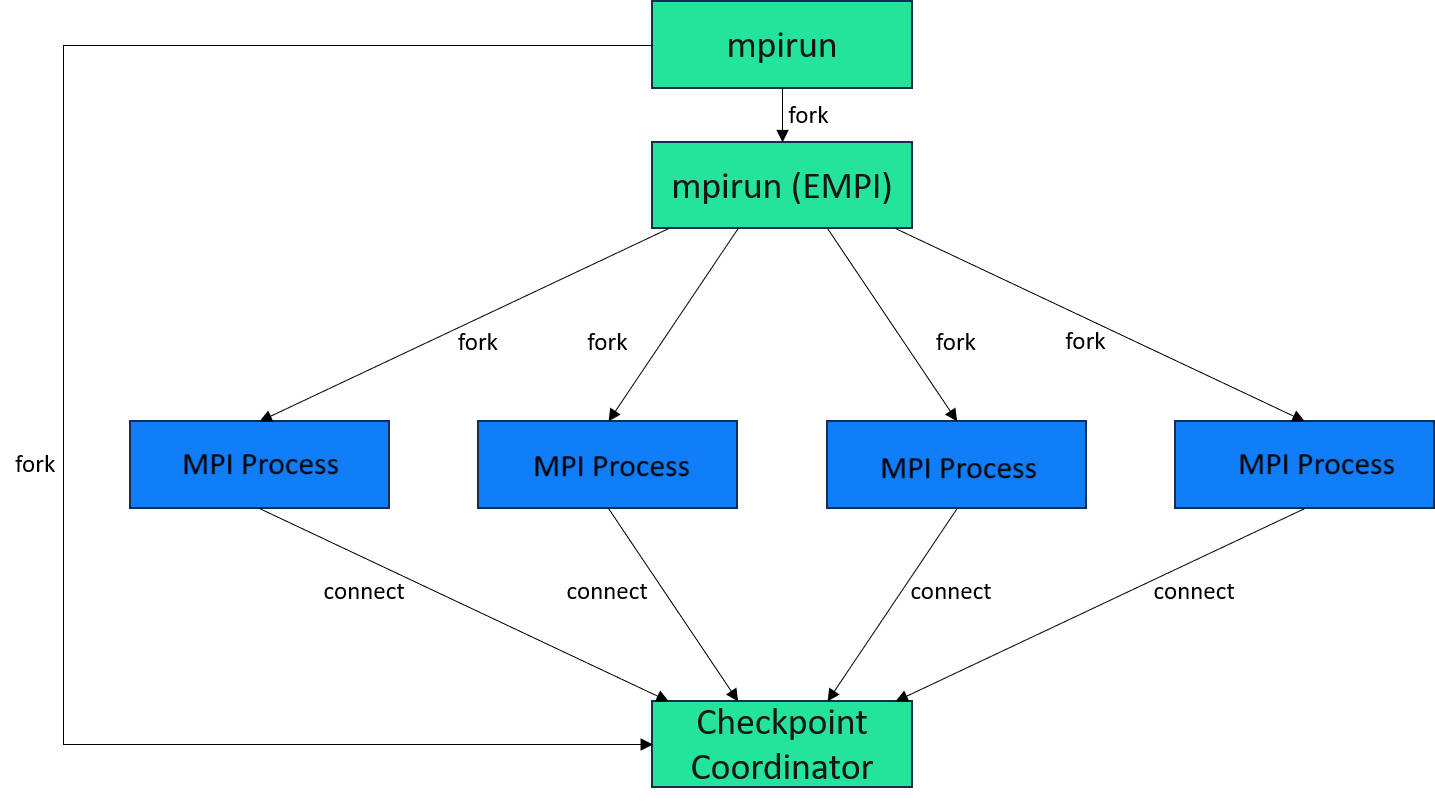}
    \caption{Process structure in each node}
    \label{fig:CoordSetup}
\end{figure}

We first set up the MPI environment and connect the MPI processes with the checkpoint coordinators. Most MPI libraries establish the MPI environment using the mpirun command. This command creates a server process, which forks child processes in which the program is executed. The server process is responsible for keeping track of process states, dynamically spawning new processes, printing the stdout output stream for all child processes in a single terminal, and performing other coordination operations. The mpirun binary provided by our library launches the external (native) MPI's (EMPI) mpirun and also launches a checkpoint coordinator program. The external MPI's mpirun forks the child processes as usual. When MPI\_Init is called by the child processes in a node, we first establish a connection between these child processes and the checkpoint coordinator of the node. This is repeated for the external MPI's server processes in each node. The process structure in each node is depicted in Figure \ref{fig:CoordSetup}.

\subsubsection{Ensuring Safe State of the Processes for Checkpoint Operations}

\begin{figure}
    \centering
    \includegraphics[scale=0.1]{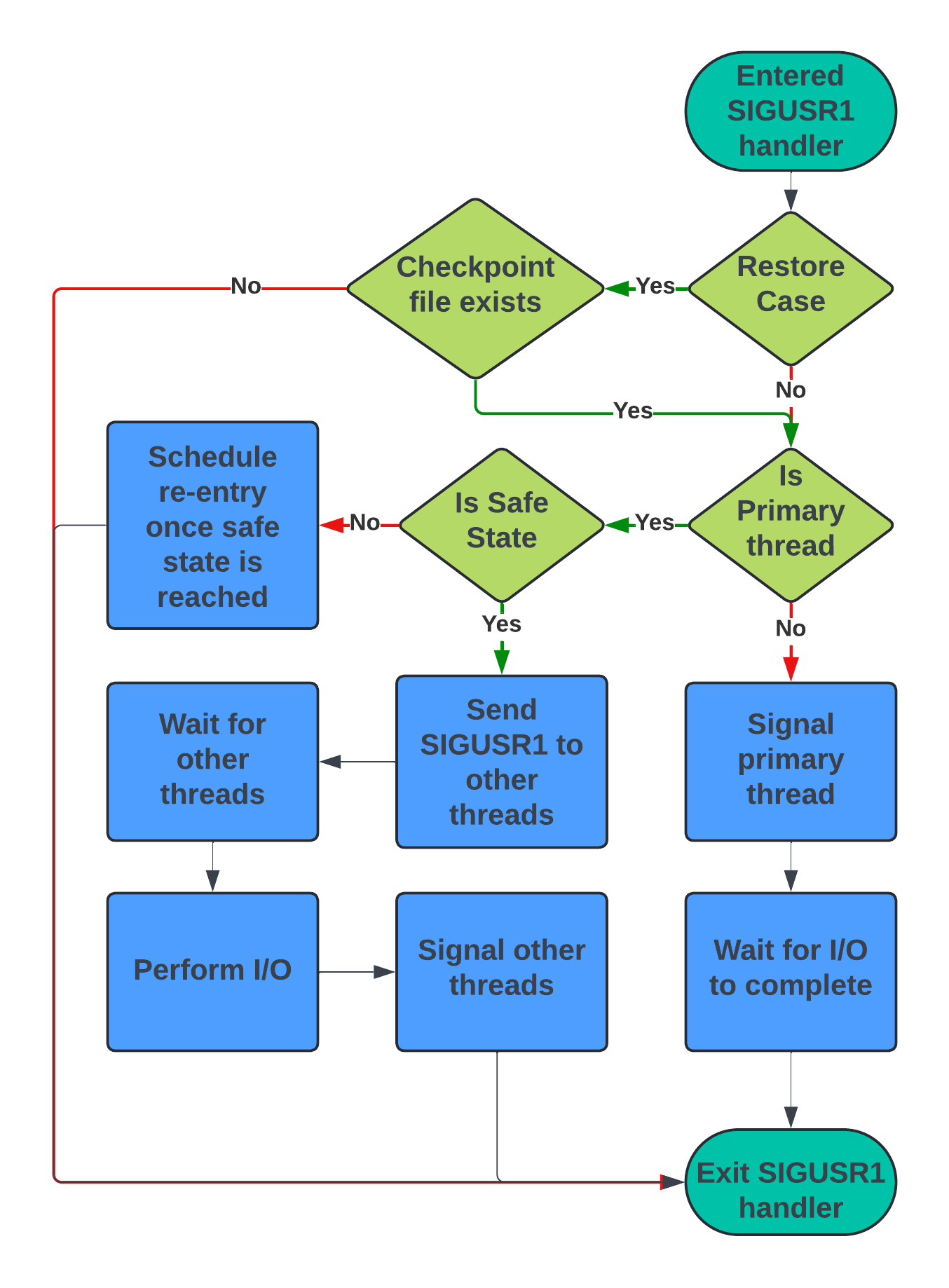}
    \caption{Signal handling procedure for SIGUSR1}
    \label{fig:SigHand}
\end{figure}

Before performing a checkpoint writing or reading operation, we need to ensure that the processes and their threads reach a safe state for the checkpoint operations. Our checkpoint coordinator periodically sends the SIGUSR1 signal to the child processes after sleeping for a certain time. Upon the receipt of SIGUSR1, the primary thread enters a signal handler and operates as described in Figure \ref{fig:SigHand}. First, we check if the primary thread is in a safe state, i.e., if it is not within an external MPI function call. If the primary thread is not in a safe state, then we return from the handler immediately but schedule to enter it again once it reaches a safe state. Once the primary threads of all processes enter the signal handler in a safe state, each primary thread sends a SIGUSR1 to all the other running threads. At this point, all the threads enter the signal handler with the primary thread waiting for them, following which writing or reading from the checkpoint file can begin.

\subsubsection{Checkpoint I/O}

Our library uses the primary thread to read/write from the checkpoint file. We cannot directly perform I/O with the checkpoint file, as it can corrupt the current stack, which will be used to perform those operations. Therefore, for reading or writing, we have to use a temporary stack that is not checkpointed itself. We use the setjmp-longjmp set of functions to jump across the original and temporary stack so that we can perform the I/O in one context without corrupting the other.
At this point, the primary thread begins I/O with the checkpoint file while the other threads wait for its completion. Each process reads or writes to a separate file. All of the I/O at this stage is performed using primitive system calls made using direct assembly code, as we do not want the libc library mappings to be modified during this operation.

\subsubsection{Checkpoint Writing}

When writing to the checkpoint file, we first write the program counter, stack pointer, and frame pointer from the originally saved execution context to the file for each thread. Following that, we write the data from segment registers and the process information structure, and then dump each memory mapping of the process to the checkpoint file, preceded by its corresponding information structure.

\subsubsection{Checkpoint Restore}
We need to restore the memory mappings in the exact same addresses as they were in the checkpointed process. This is because those mappings together form a unique address space with a complex web of pointers pointing across those addresses. If even one mapping is at a different address than its original, all the pointers to it from other mappings will now be pointing at an invalid address, which could eventually lead to segmentation faults. For the restore operation, we first read from the checkpoint file up to the process information structure. We modify the brk value to match that of the checkpointed process (obtained from the process information structure). Following that, we unmap all of the current memory regions from the saved array of structures of the memory mapping information, except the mappings of our library (where the current code would be running from). At this point, only the temporary stack and our library's mappings would exist in the current process. We then read the memory mapping information from the checkpoint file, create a new mapping at the same address, and read the entire mapping dump into that new mapping. This is repeated for every mapping dumped into the checkpoint file. Therefore, in the end, we would have an exact copy of the original process with the same address space fully restored within the new process.

\begin{figure}
    \centering
    \includegraphics[width=\linewidth]{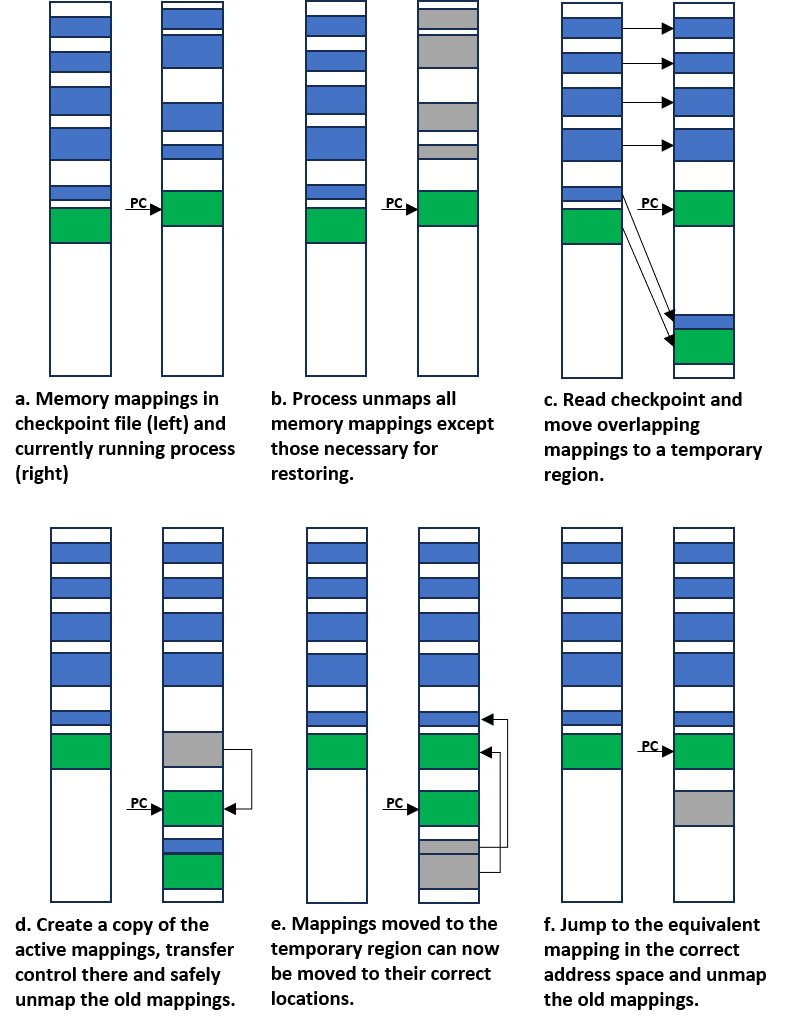}
    \caption{Procedure for restoring memory mappings}
    \label{fig:RestoreMap}
\end{figure}

However, as we scale up to many processes, we can encounter exceptional cases, like when one of the mappings that needs to be read overlaps with our library's mappings or the temporary stack. While this case is extremely rare, it can still occur frequently in at least one out of thousands of processes. Tools like DMTCP tend to abort when these cases are encountered. However, as we will test this library in heavy failure conditions where any process, both old and new, can fail at any time, and failures can occur 5-20 times in the whole execution, all these cases must be accounted for. When such a case of overlapping memory mappings is encountered, these mappings are instead mapped at a safe address that is free in both the checkpointed and restoring processes and later remapped to their original address space.
This procedure of restoring the original address space is described in Figure \ref{fig:RestoreMap}.

\subsubsection{Completion of the Checkpoint Operations}
We can now use longjmp with the original execution context to get back to the original state.
We now unmap the temporary stack mapping, and for the restore case, we also unmap the latest mappings of our library, which we had not unmapped yet. This is safe as the currently running code would exist in our library's mapping that was restored from the checkpoint file. We then synchronize across the processes using a barrier operation that is conducted using the coordinators, recover any lost messages for the restore case, update the file denoting the latest checkpoint for the checkpoint creation case (covered in more detail in later sections), signal the coordinator to restart the checkpoint timer, and return from the signal handler after releasing all the acquired locks.

\subsection{Replication}

The replica of a process can be defined as a process that performs the same operations in the same order on the same inputs and produces the same outputs at the application level. We denote the process that has a replica as the {\it original} process and the process that replicates the original process as its {\it replica} process.
Our library creates replica processes at the application level.
The state of the application can be defined by the data, heap, and stack segments. Therefore, these are the only segments we need to copy from an original process, A, to another process, A', such that A' can become the replica of A. However, the primary challenge is the difference between the address spaces of A and A'.
The heap and stack segments are generally located at different virtual address spaces in different processes. Therefore, we cannot copy the data to the same address in the destination, as it would likely be unmapped or mapped by a different segment. We also cannot simply move just the data of the heap and stack segments into a different address space. There could be many pointers that exist in the heap and stack that would be pointing to addresses in their original address space. We also cannot iteratively modify all of these pointers by applying appropriate offsets, as there is no guarantee of which of these values are pointers and which are just large numbers. There is also no guarantee of the pointers being aligned to certain addresses.

Our solution to this issue involves creating a new heap and stack that is located at a common address space across all the processes. Inside MPI\_Init, after connecting to the checkpoint coordinator and initializing the external MPI's MPI\_Init, all of our processes communicate to identify two contiguous regions of memory that are unmapped across all the processes. We create two mappings in these regions. One acts as an {\it application space heap}, and the other acts as a {\it common address stack}. All allocation-related functions like malloc, free, etc., use the regular heap when called from inside our library, but use the application space heap when called from application control.
After the two mappings are created, we move the stack pointer to the common address stack and move the program counter to the start of the main function using setjmp and longjmp. Through this, the stack is rebuilt from the start of the main function in the new common address stack, and the older stack is never used again until MPI\_Finalize is called.

On the second entry to MPI\_Init, this time from the common address stack, all of our processes are ready to be replicated as they now hold all the relevant allocations in the application space heap and have a fully functional stack built up in the common address stack. Both of these mappings are at the same virtual addresses across all the processes, and any pointers filled within them have also been obtained from that common address space. Therefore, we can now copy the data, heap, and stack without any invalid addresses. Our library supports partial replication. The user can define the degree of replication using mpirun arguments. Let us consider that the user wants to execute the application on $N$ processes with $M$ ($M \leq N$) of the $N$ processes to be replicated. Hence, the application is started with a total of $N+M$ processes. The first $N$ processes with ranks $0$ to $N-1$ are treated as the $N$ {\it original} processes, and $M$ of these $N$ original processes will have replicas. Each of the remaining $M$ processes, from ranks $N$ to $N+M-1$, is mapped to be a replica of a distinct original process, ranked from $0$ to $M-1$, and is prepared to receive the replication data from its original process. We also perform this copy from within a temporary stack mapping, just like checkpoint/restart, to ensure that the stack segment is not disturbed during the copy operation. For this work, we have shown results and benefits with the full replication case ($N=M$).

\begin{figure}
    \centering
    \includegraphics[width=\linewidth]{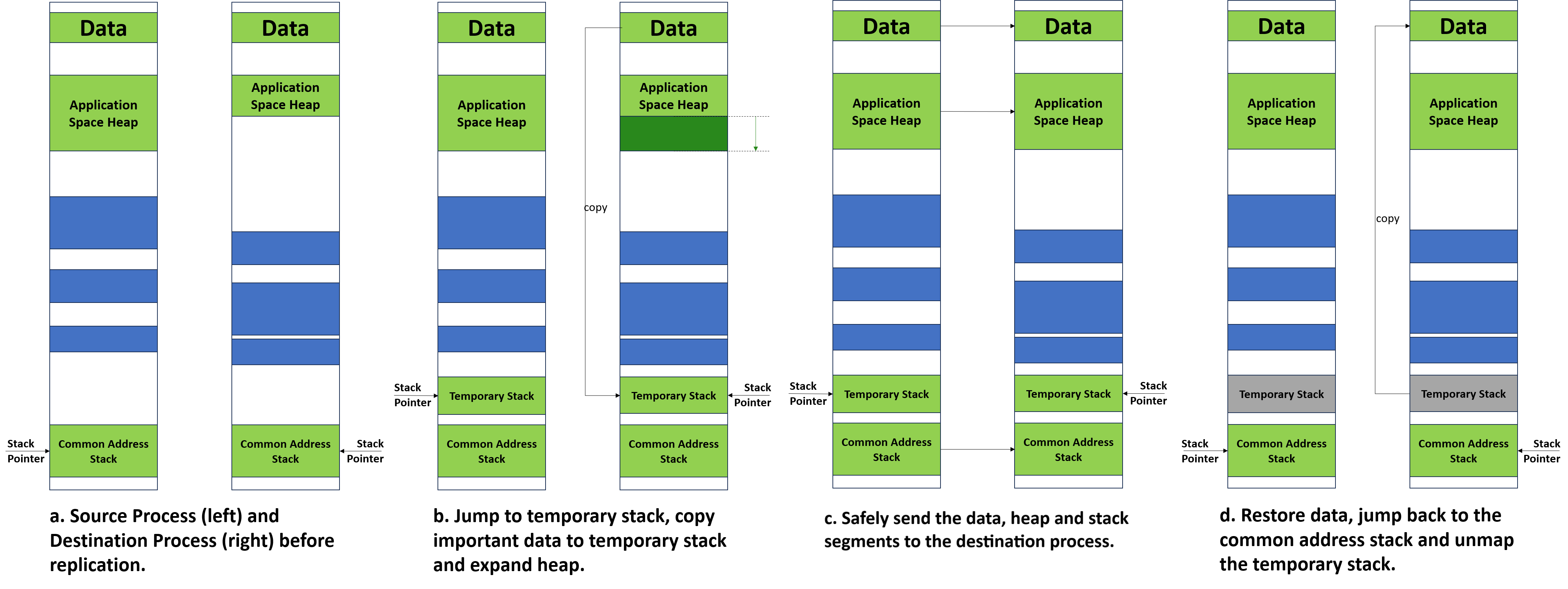}
    \caption{Procedure for replicating a process}
    \label{fig:ReplicationProcedure}
\end{figure}

The procedure for replicating a process is described in Figure \ref{fig:ReplicationProcedure}. Before copying the data segment, we save some important structures, such as MPI communicators and datatypes defined in the data segment, which need to be preserved as unique entities in the temporary stack. These are restored to their original values once the data segment is restored. We expand or contract the heap segment to match the size of the original process before copying it over. Furthermore, as this application space heap is managed by a custom allocator, we only need to copy the memory regions that have been marked as allocated. For the stack segment, we use the stack unwinding functionality provided by the {\it libunwind} library to store the values of the pointers used in each stack frame, copy them over to the destination before copying the stack segment, and reset these pointer values to the correct ones at the destination process. As an example, the stack frame may hold pointers to certain libraries like libc in order to restore them into caller-saved registers upon return from a function. These pointers need to be fixed to point to the same library in the replica process. Note that these are very rare cases of limited and specific locations in each stack frame that need to be modified and are only used for maintaining register values across functions. We also send pointers to the application space heap's free lists, heads of message logs, and non-blocking communication request lists that we maintain, which are specifically allocated in the application space heap even though they are allocated from within our library's code, as we want that information to be copied to the replica. Finally, we can synchronize all our processes and return from MPI\_Init with some processes acting as replicas of unique processes.

\subsection{Integrating Checkpoint/Restart and Replication}
\label{unifying}

There are a few nuances involved in using the checkpoint/restart and replication mechanisms in tandem. Firstly, process failures will no longer result in the job immediately restarting as long as one replica for each unique process is alive, but the total number of live processes will reduce. Therefore, subsequent checkpoints would need to be made with a lesser number of processes. Furthermore, when process failures result in the failure of the application itself, e.g., due to the failure of both the original and its replica process, and the application is restarted, the number of processes with which it is restarted can be different (either more or fewer) from the number of processes used to create the last checkpoint. This is because our FTHP-MPI fault-tolerant framework is flexible and supports partial replication. In cases when the application is restarted with $N+M$ processes, i.e., the same number of processes the application was started with, the last checkpoint is likely to have been created with fewer than $N+M$ processes due to the application continuing execution in spite of failures of some of the processes. In some other cases, spare processors may not be available to replace the failed processes, resulting in the application restarting with a lesser number of processes with a lower replication degree (partial replication) than the number of processes used for the last checkpoint. In both cases, the job will need to be able to fully restore to the previous state.

To resolve this, we perform incremental checkpointing by using the idea of baseline checkpoints that both address the above-mentioned issue and also lower our checkpoint/restart overhead. When MPI\_Init is called, before calling the external MPI's MPI\_Init, we have all the processes, both original and replica processes, create a checkpoint that we call the baseline checkpoint. From then on, all of the future checkpoints are not full memory dumps but only the data that would be transferred over for replication (data, heap, stack, and some other pointers) and are only created by the original processes. Upon application restart, each of the original and replica processes first reads from their baseline checkpoint files, then initializes the external MPI's MPI\_Init and creates all the required communicators, and then the original and its corresponding replica process read from the latest checkpoint files. The read from the baseline checkpoint ensures that all the processes have the same addresses as the application space heap and common address stack as the original set of processes. The following reads from the latest checkpoint essentially updates the data, heap, and stack segments to advance the application state to its latest saved state. Reading from the baseline checkpoint allows the replica process to safely read from the same latest checkpoint file as its original counterpart, as recovering the original address space guarantees that the common heap and stack addresses will not overlap with any of the existing mappings in the replica process.
\section{Interfacing with the External MPI Library}
\label{using-multi}

Production supercomputing systems have native MPI libraries\footnote{This paper refers to native MPI libraries interchangeably as external MPI libraries.} that are specifically tuned to exploit the underlying hardware architecture, including the network topology (e.g., Dragonfly topology in Cray systems), to maximize performance.
Generic MPI libraries typically do not take advantage of the underlying hardware architecture since they follow generic communication algorithms, which makes them less desirable. Therefore, users are heavily dependent on support from the native MPI libraries in order to enable fault tolerance without performance loss. We have designed our library in a way that allows the user to utilize the efficient implementations of the native MPI libraries while also having replication-based fault tolerance without any changes to the code bases of the native MPI libraries. Our approach involves dynamically loading the native MPI library at runtime. Our library sits on top of this native or external MPI library as a wrapper and calls the appropriate underlying functions based on the user's request, while also involving replicas in communications as needed.

\subsection{Wrapping around the External MPI Library}

Our library uses structures that act as wrappers for all of the standard MPI handles, such as MPI\_Comm, MPI\_Request, etc. The elements of these structures are collections of one or more corresponding handles from the external MPI library and a few other communication-related variables. We use a script to extract all the "\#define", "typedef", and "enum" declarations in the mpi.h file in the external MPI library and modify them such that all the instances of the pattern MPI are replaced with EMPI to refer to the external MPI library. Furthermore, all MPI functions are dynamically loaded at the start of MPI\_Init using the functionality provided by the dl library in the UNIX API. With this, the functions and symbols of the external library can be used without conflicts by simply using EMPI keywords for all the calls.

\subsection{Hiding Failures from the External MPI Library}

The mpirun server processes in MPI libraries use system calls like poll and waitpid to check for process failures. Upon failure detection, libraries without built-in fault tolerance generally proceed to kill all the child processes. We cannot allow this to happen, as it would invalidate our main objective of tolerating failures until all replicas of a process are dead. We, therefore, preload the external MPI's mpirun with a small proxy library when it is called from within our mpirun program. This proxy library intercepts calls to functions like poll and waitpid, stores their original outputs, and modifies the outputs before returning them to the user, such that all process failure incidents are hidden from the external MPI's server process. The original outputs are only returned in the same order as they were obtained once the program either finishes or aborts due to all replicas of a process dying.

\section{FTHP-MPI Library Implementation}
\label{library}

Our MPI library uses the external MPI library for communications. We start more processes than required by the user and convert the extra processes into replicas. Internally, we classify the processes launched in the MPI job as a computational process and a replica process for our implementation purposes. This is defined by the communicators to which the processes belong. Our library provides all the standard MPI handles as pointers to internally defined structures. These structures hold one or more of the external MPI library's corresponding handles. For example, the structure pointed to by MPI\_Comm holds six external MPI\_Comm elements.
We use these six communicators in our implementation for the external MPI library for communication:\\
\begin{enumerate}
    \item \textbf{eworldComm}: This is initially just a duplicate of the external MPI\_COMM\_WORLD.
    \item \textbf{EMPI\_COMM\_CMP}: This communicator contains all the computational processes.
    \item \textbf{EMPI\_COMM\_REP}: This communicator contains all the replica processes.
    \item \textbf{EMPI\_CMP\_REP\_INTERCOMM}: This is an inter-communicator bridging the communicators for computational and replica processes, respectively.
    \item \textbf{EMPI\_CMP\_NO\_REP}: This communicator contains all the computational processes that do not have a replica.
    \item \textbf{EMPI\_CMP\_NO\_REP\_INTERCOMM}: This is an inter-communicator bridging the EMPI\_CMP\_NO\_REP communicator with the replica communicator.
\end{enumerate}
All of these internal EMPI communicators are initialized from MPI\_Init and modified whenever a failure occurs. Therefore, the user can use the same handles provided by our library in the same way as a standard MPI library without changes to their code, while all the failure handling is performed internally.

The replica processes in our library perform the same computation operations redundantly along with their corresponding computational processes. They need to participate in the communications to obtain the latest data as well. Therefore, they must identify the proper source and/or destination for any communication. Our library performs efficient communications that involve replicas by parallelizing these communications as much as possible. In general, any communication can be defined as a data transfer from a set of source processes to a set of destination processes. A subset of the set of source processes and a subset of the set of destination processes may have replica processes. In our library, all the computational source processes communicate with their replica destination processes as normal, using the communicator holding all computational processes. All the replica source processes communicate with their replica destination processes using the communicator holding all replica processes. When there is a replica destination process for which the corresponding replica source process does not exist, then the computational source process acts as a source for that replica process using one of the inter-communicators as needed. In the case where there is a source replica process but the corresponding destination replica process does not exist, the communication can be skipped in the source replica process. This setup ensures that any message from a specific source computational and replica process arrives in the same order in the destination computational and replica process, respectively. However, due to network fluctuations, messages from different source processes can arrive in a different order in a computational process as compared to its replica. In programs where wildcard communications involving MPI\_ANY\_SOURCE are used and the order of the received messages changes the final output, we need to ensure that the received messages are returned to the user in the same order in both the computational and replica processes. Therefore, for such communications, after the computational destination process receives a message from a source, it sends the rank of the source process and the tag of that message to its replica. The replica destination process waits for this information from its corresponding computational process for such communications and then receives the actual message from the corresponding source computational or replica process (if it exists) with the same tag. This ensures that the messages are returned to the user in the same order across the two processes, even if they arrive in a different order.
All these communications are performed in parallel and ensure that the correct data is received in the correct location in both the computational and replica processes.

\begin{figure*}
    \centering
    \includegraphics[width=\textwidth]{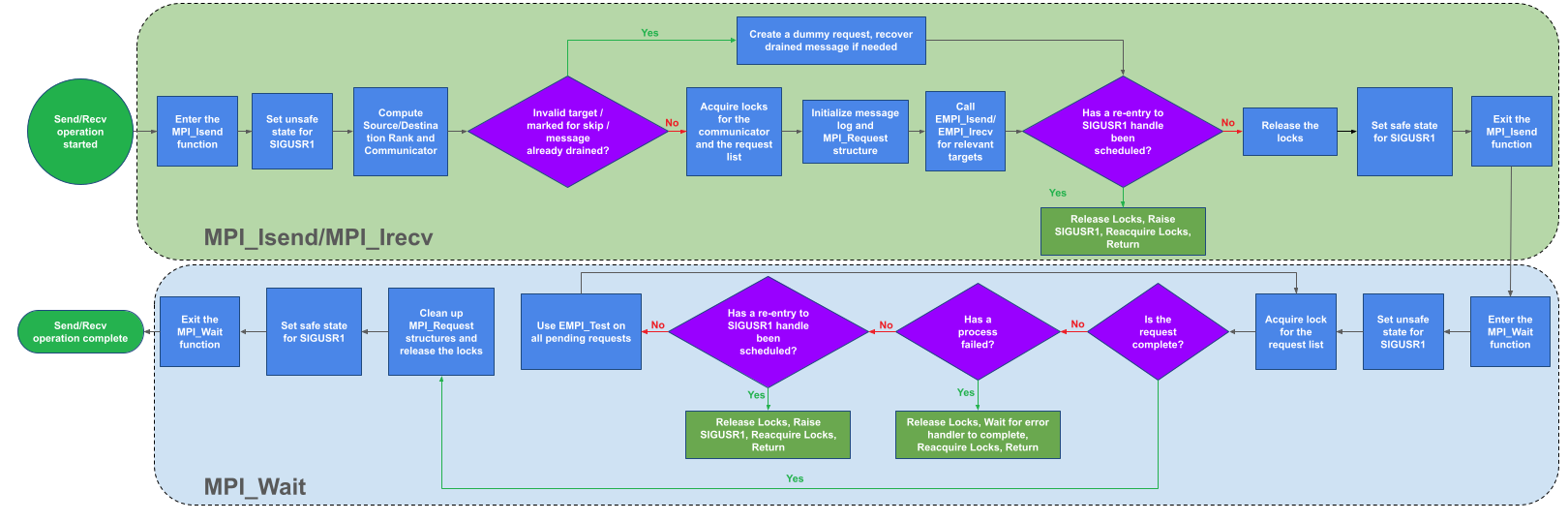}
    \caption{Communication Workflow}
    \label{fig:CommFlow}
\end{figure*}

All of our communication implementations utilize non-blocking EMPI functions like EMPI\_Isend and EMPI\_Irecv. Since we want to safely checkpoint the application and also expeditiously handle any failures that may arise, we never use blocking EMPI functions. The blocking MPI functions provided by our library, like MPI\_Recv, internally call their non-blocking EMPI variants like MPI\_Irecv followed by MPI\_Wait. The communication workflow is depicted in Figure \ref{fig:CommFlow}. Furthermore, we have also implemented collective communications in a similar manner using the non-blocking communications like EMPI\_Ibcast and EMPI\_Iallgather. We use a similar approach to peer-to-peer communications to parallelize these communications across the computational and replica processes. When there are computational processes without replicas that provide input data to the collective, they also propagate that data to the appropriate replica processes using one or more parallel collectives. This allows us to use the efficient implementations of the collective communications from the EMPI library.
\section{Failure Management}
\label{failure-mgmt}
Our library is designed to handle fail-stop errors. These are errors in which one or more running processes fail due to some arbitrary issue that could have a variety of causes, both at the hardware and software levels. We have mechanisms that detect these failed processes, propagate that information to all the remaining processes, and modify the internal structures accordingly.

\subsection{Failure Detection and Propagation}
We had mentioned in Section \ref{using-multi} that we intercept the system calls used by the external MPI's server process to hide the failures in the external MPI environment. This allowed us to continue running the application as the external MPI library would not forcibly abort it. Another advantage of that interception interface is that we can also identify when a process has died. When a system call like waitpid is called by the external MPI library's server process, it is intercepted by our proxy library. At each failure, the proxy library also establishes a connection to the coordinator running on its node and informs it about the failure. For system calls like waitpid, the PID of the failed process is explicitly sent. For system calls like poll, there is no way for our proxy library to identify which specific processes have failed since it uses file descriptors as input. Therefore, we instead signal to the checkpoint coordinator that some miscellaneous process has failed, which is then verified by the coordinator by polling all the local MPI processes. The coordinators propagate the information about the failed processes among themselves and then propagate it to the MPI processes.

\subsection {Repairing the World}

\begin{figure}
    \centering
    \includegraphics[width=\linewidth]{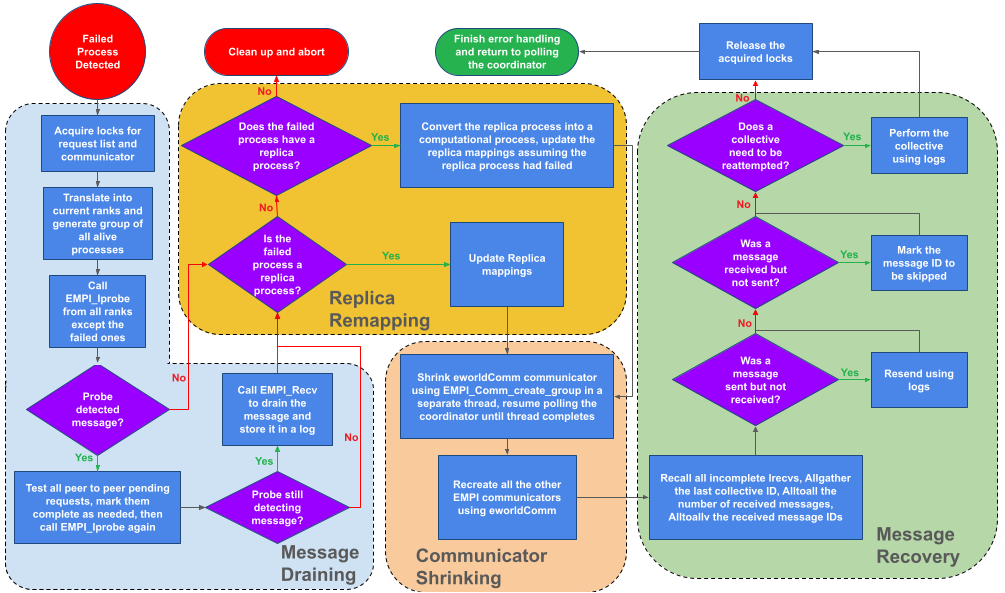}
    \caption{Workflow for handling failures}
    \label{fig:FailFlow}
\end{figure}

Once all the MPI processes receive information about the failed processes from their local coordinators, they first attempt to drain any in-flight messages and then repair the MPI environment. This is depicted in Figure \ref{fig:FailFlow}. We use the communicator shrinking mechanism defined in ULFM for this purpose. Communicator shrinking involves removing all the failed processes from the communicator. This is done using EMPI\_Comm\_create\_group on the communicator holding all the MPI processes. This only requires communication between processes in the smaller communicator. If the failed process is a replica process, it is dropped, and the ranks of some replica processes and the computational replica maps are updated. If the failed process is a computational process that has a replica, then the newly shrunk communicator has its processes shuffled such that the replica now becomes the computational process, following which it is considered that the replica was the one that had failed. Converting to a computational process at this point simply involves creating the correct set of EMPI communicators that are regenerated using the shrunk processes.
Once the communicator holding all the processes is shrunk, all the other communicators can be recreated using it. Unlike the ULFM standard, we do not depend on returning error codes to the user. Our entire shrinking mechanism is conducted on a separate thread, and thus, no process waits for every other process to enter an MPI function before shrinking can begin. This is a big advantage when dealing with embarrassingly parallel applications or applications in which certain processes do not enter MPI functions for a long period of time.

\subsection{Message Recovery}
\label{msgrecovery}
With the communicators appropriately repaired, the final task is to account for any transmissions lost during the repair process. This is handled using the logs that we maintain during the communications. We use a sender-based message-logging mechanism where the sent data is saved on the sender side, and a send ID is piggybacked on every message. When a failed process is detected, before we shrink the communicator, we drain all the processes of any messages that may have been sent to them. After repairing the communicators, all the messages sent by a process but not received at the destination are resent using the message logs, and all the messages received by a process but not sent from the source are marked using their send IDs to be skipped in the future. An example of this could happen when a replica process converts into a computational process, and the replica process may already have received certain data from other replica processes that its computational process would not have received before failure.
\section{Experiments and Results}
\label{exp-res}
We performed our measurements on a large-scale system consisting of 300 compute nodes with 48 cores per node, 4GB DDR4 RAM per core, and an Infiniband interconnect. All our checkpoints were saved to a Lustre parallel file system mounted using POSIX, with 4PB of global storage and a throughput of 100GB per second. We scaled our experiments up to 8192 processors. All our experiments have been conducted using the MVAPICH2 library as the underlying native/external MPI library. Our experiments with replication use dual redundancy in which 50\% of the processors are used for replication. Note that our replication-based framework and experiments do not allocate twice as many processors as in experiments without replication. We argue that, at a high MTBF, given a fixed number of processes, it is better to use half of them as replicas rather than all of them as computational processes.

We used the High-Performance Conjugate Gradient (HPCG) Benchmark, the CloverLeaf mini-application\cite{mallinson-cloverleaf:-2013}, and the Plasma Particle-In-Cell (PIC) simulation skeleton codes\cite{decyk-skeleton-1995} to test our library. HPCG is a C++ based, weak-scaling benchmark that measures the performance of basic operations like sparse matrix-vector multiplication in a unified code. The benchmark first runs a sample iteration to obtain its execution time, by which it computes the number of iterations that will be needed to approximately achieve a target runtime given as input. As the number of processes increases, the target time remains the same, but the total data being operated upon increases, with each processor having its own local data. 
The PIC simulation skeleton codes simulate the movement of charged particles in an electromagnetic field that they themselves produce. This FORTRAN application divides the field into a grid distributed across processors and consists of a series of iterations in which the charge on each grid cell is accumulated to obtain source densities, which are used to compute the resulting electromagnetic field, which is then used to compute the movement of the particles across the grid.
CloverLeaf is a FORTRAN-based mini-application that solves the compressible Euler equations on a Cartesian grid using an explicit, second-order accurate method. It operates on a system of three partial differential equations for the conservation of mass, energy, and momentum, respectively. The equations are solved on a staggered grid in which each cell center stores three quantities, namely, energy, density, and pressure, and each node stores a velocity vector. In all the cases, the optimal checkpointing interval has been computed using the Young-Daly formula using the MTBF and checkpoint creation time and has been recorded in Table \ref{tab:ckptints}.
\begin{table}[ht]
\centering
\resizebox{\linewidth}{!}{%
\begin{tabular}{|c|c|c|c|c|}
 \hline
 Application & Number of  & MTBF ($\mu$) & Checkpoint Creation & Optimal Checkpointing \\
 & Processes & (seconds) & Time (C) (seconds) & Interval ($\sqrt{2\mu C}$) (seconds) \\
 \hline
 HPCG & 1024 & 16000 & 46 & 1213.26 \\
 HPCG & 2048 & 8000 & 65 & 1019.80 \\
 HPCG & 4096 & 4000 & 114 & 954.98 \\
 HPCG & 8192 & 2000 & 215 & 927.36 \\
 CloverLeaf & 2048 & 2000 & 44 & 419.52 \\
 CloverLeaf & 4096 & 1000 & 45 & 300 \\
 CloverLeaf & 8192 & 500 & 42 & 204.93 \\
 PIC & 2048 & 2000 & 66 & 513.81 \\
 PIC & 4096 & 1000 & 63 & 354.96 \\
 PIC & 8192 & 500 & 60 & 244.94 \\
 \hline
\end{tabular}
}
\caption{Optimal Checkpoint Intervals}
\label{tab:ckptints}
\end{table}

We show results of HPCG in terms of both performance measured in FLOPS (floating point operations per second) and application efficiency. In our work, we have defined application efficiency of an execution as the ratio of the performance (in FLOPS) per core of that execution to the performance (in FLOPS) per core of the failure-free execution at 1024 processes. Under failure conditions, the program will incur additional time or redundancy-based overheads, which will lead to a lower performance and thus, a lower efficiency for that scale. For checkpoint/restart, efficiency will also be lost due to the checkpointing overhead. Replication, on the other hand, would incur a direct 50\% efficiency loss simply by using the framework, as half the processes would just do redundant work.

We first show results with HPCG. We have chosen the HPCG target execution time as 3 hours to meet our job time limit constraints across a large number of executions. We used a failure simulator to kill random processes with MTBF set to 2000 seconds at 8192 processors, which is sufficient to have a noticeable impact given the benchmark's runtime, as higher MTBF values require longer-running executions to be measured fairly. We expect our results to hold similar patterns when a longer-running application is used with a higher MTBF value. Since the MTBF value doubles when the number of processors is halved, we set the MTBF to 4000 seconds for 4096-processor executions and so on. Our failure simulator killed processes at time intervals based on a Weibull distribution of shape 0.7, which was shown to closely match the failure distribution in real systems\cite{schroeder-large-2009}.

We compared the performance and efficiency using checkpointing and full replication as the number of processes scales up, where each process is mapped to a processor core. For fair comparison, we have used the same number of processor cores for both checkpointing and replication, including the processor cores used for running the replica processes. For example, 8192 processes with checkpointing use all 8192 processes for their computations, while the replication case only uses 4096 processes, and the other half does the same computations redundantly. As the number of processes is the same, both cases are also simulated with the same failure rates. As we assign the latter half of our process set to replicas, replica processes generally run on different nodes than computational processes. The 1024 process execution without any simulated failures is used as the baseline for efficiency calculation. These metrics have been obtained over an average of five runs.

\begin{figure}
    \centering
    \includegraphics[width=\linewidth]{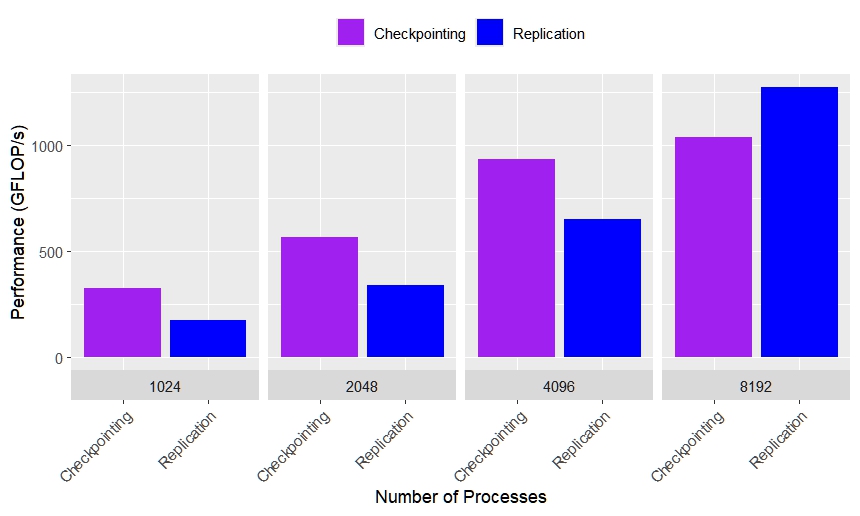}
    \caption{HPCG Performance}
    \label{fig:perfHPCG}
\end{figure}

\begin{figure}
    \centering
    \includegraphics[width=\linewidth]{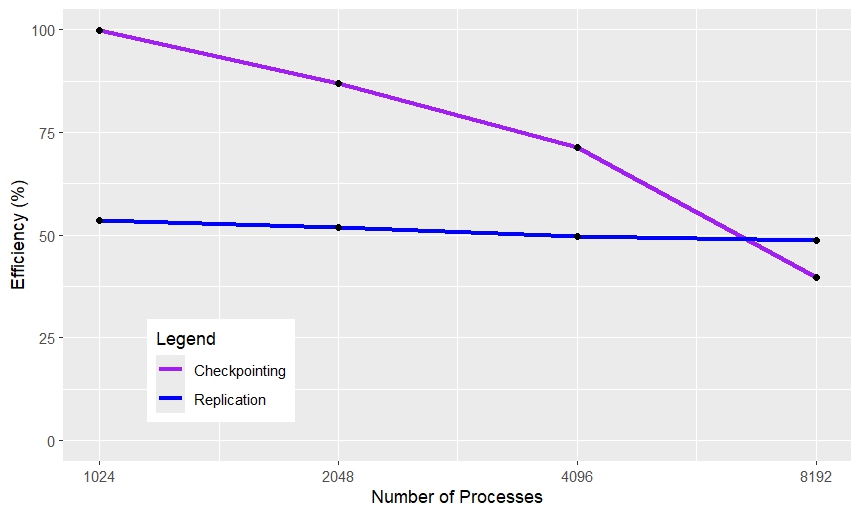}
    \caption{HPCG Efficiency}
    \label{fig:effHPCG}
\end{figure}

Figures \ref{fig:perfHPCG} and \ref{fig:effHPCG} show the comparisons in terms of FLOPS and efficiency, respectively. Here, we can observe that at 1024 processes, the failure rate is so low that the checkpointing case incurs very little cost and achieves near 100\% efficiency. On the other hand, the replication case immediately loses around 50\% efficiency due to 512 processes out of the 1024 processes performing redundant work. However, as the number of processes scales up and the failure rate increases, we observe that the checkpointing case experiences a significant efficiency drop. On the other hand, the replication case incurs a near-negligible efficiency drop, and its performance nearly doubles when the number of processes doubles.

As the total amount of data increases with the number of processes, the checkpoint time also increases. Therefore, the checkpoint/restart framework would lose efficiency very rapidly as it scales up due to a combination of checkpoint and failure overheads. However, as the replication approach has a much lower failure overhead and no checkpoint overhead, the efficiency would drop at a much lower rate as the number of processes scales up.
This culminates in the replication case providing 18.18\% higher performance than the checkpointing case at 8192 processes, even though half of its processes are doing redundant work, as the efficiency for checkpointing drops to below 50\% due to heavy checkpointing and failure overheads. With MTBFs decreasing with an increase in the number of processes, we reach the threshold of failure rates where replication would outperform pure checkpoint/restart.

The two figures together show that while the application scales with the addition of more processors for both checkpointing and replication, better efficiency is obtained beyond a certain number of processors (in this case, 8192 cores) by using the additional processors for replication rather than using all the processors for application execution with checkpointing. 
These are significant results that show the benefits of only replication over checkpointing using actual executions and match the efficiency patterns in the simulations by Ferreira et al. \cite{ferreira-evaluating-2011}.
Similar to this work, our work shows increased efficiency with replication over checkpointing (about 3000 seconds of MTBFs in their work versus 2000 seconds in our work). However, while the earlier work shows the benefits of replication combined with checkpointing, our results show that for small execution times, replication alone is sufficient.

For a fixed MTBF, this pattern would persist until execution time exceeds a threshold, at which point the probability of both replicas of a process failing will be high, and to the extent that replication with a long checkpoint interval would be more suitable than pure replication. In all our experiments involving relatively small execution times, we did not encounter a case where both a computation and its replication process failed, resulting in the application failure, and hence the need for checkpointing.

\begin{figure}
    \centering
    \includegraphics[width=\linewidth]{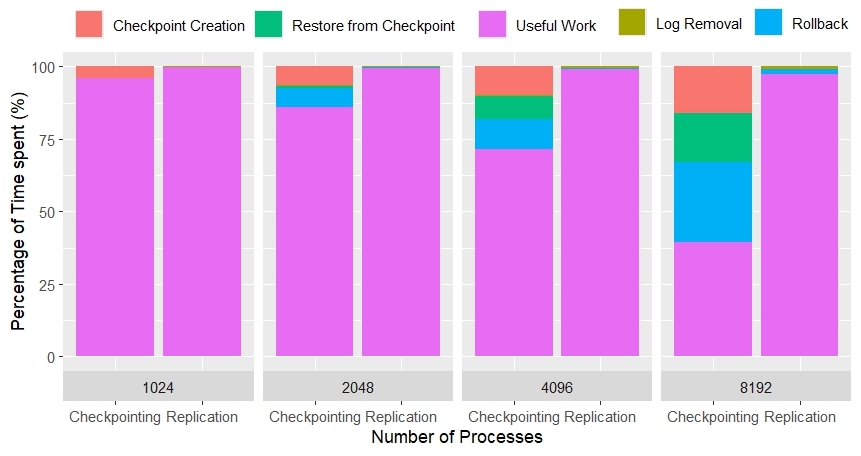}
    \caption{Time distribution}
    \label{fig:tdistHPCG}
\end{figure}

The overheads due to checkpointing and replication under failure conditions are further analyzed in Figure \ref{fig:tdistHPCG}. The rollback component involves the time lost due to regressing back to a previous checkpoint state, the time taken to recreate all the communicators, and the time taken to recover any lost messages after the communicator recreation. In checkpointing, most of the rollback time is dominated by the time lost due to regressing to the previous checkpoint. In replication, there is no regression to a previous state, and so the rollback only consists of the time taken for communicator recreation and message recovery for each failure, which is negligible. We find that the log removal component, which involves cleaning up the message logs, mentioned in Section \ref{msgrecovery}, whenever they exceed a certain memory limit, is negligible in all the cases. We can observe that replication mostly does useful work, even though half of it is redundant. Checkpointing, on the other hand, incurs more and more overhead as it scales up, resulting in less than half the time being spent on doing useful work. We also see that as the number of processes and consequently, the MTBF increases, the checkpoint creation, restore, and rollback times scale up significantly in the checkpointing case and end up occupying over half the total runtime of the application. Replication, on the other hand, incurs no checkpoint or restore overheads and only incurs a negligible rollback cost.

\begin{figure}
    \centering
    \includegraphics[width=\linewidth]{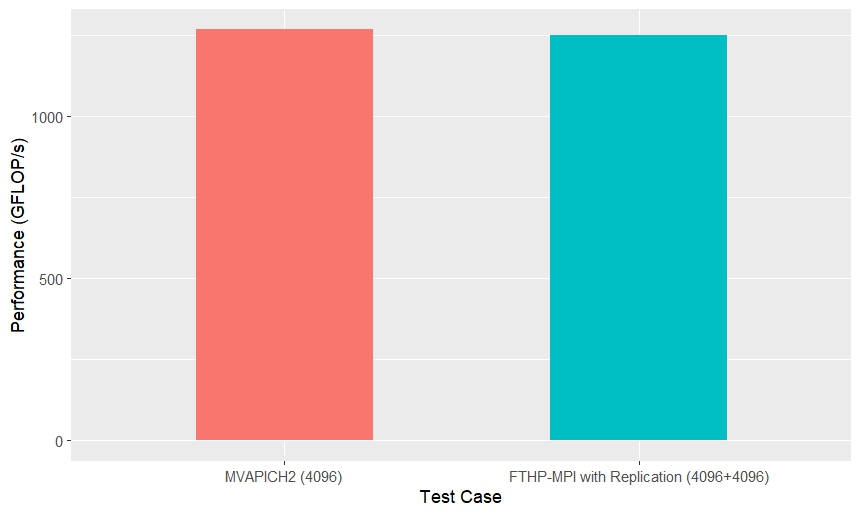}
    \caption{Failure Free experiments}
    \label{fig:failfreeHPCG}
\end{figure}

We also measured the impact of our library interceptions and the additional communications due to replicas. Figure \ref{fig:failfreeHPCG} shows the performance of the MVAPICH2 library at 4096 processes and our library at 8192 processes, with 4096 of these processes used for replication. These experiments were conducted without any failure conditions. As 8192 processes with replication are equivalent to 4096 processes, any loss incurred by our library here would be due to additional communications incurred by replicas and our library intercepting the MPI functions and other system calls. We observe that these losses are negligible, with the baseline MVAPICH2 case only showing a 1.3\% better performance than our library. Therefore, the losses due to replication almost entirely come from redundancy, and by using our library, the user incurs minimal losses compared to using MVAPICH2 directly but gets access to both checkpointing and replication frameworks for fault tolerance.

\begin{figure}
    \centering
    \includegraphics[width=\linewidth]{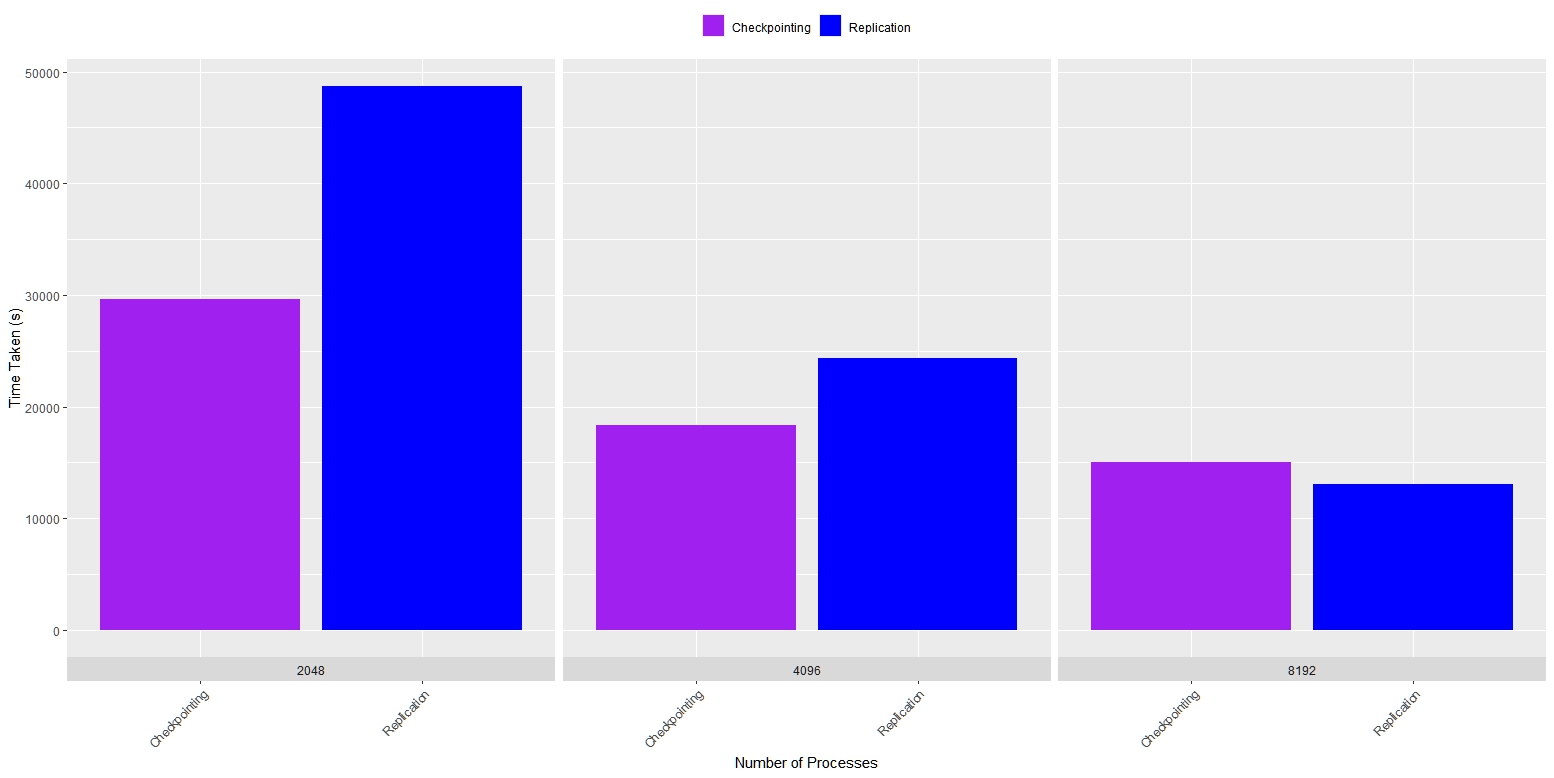}
    \caption{Comparison of Execution Times for Checkpointing and Replication for the CloverLeaf Application}
    \label{fig:cltime}
\end{figure}

We also compared the times taken by CloverLeaf and PIC scaling up to 8192 processes under failure conditions with checkpointing and replication. Here, we have used an MTBF of 500 seconds to represent more extreme failure conditions while also accounting for the smaller runtime of these applications, which require a smaller MTBF for failure conditions to be noticeable. The CloverLeaf execution was performed using the standard 8192 process benchmark dataset consisting of 122880 cells in both x and y direction and running for 2955 steps. The PIC execution was performed using the 3D Parallel Darwin Spectral code (pdpic3) with 1500 particles distributed across all x, y, and z directions and running for 1000 iterations. These timings were obtained over an average of three runs each. Figures \ref{fig:cltime} and \ref{fig:pictime} show the comparisons between the execution times for checkpointing and replication in CloverLeaf and PIC, respectively. At a lower number of processes, the checkpointing case has a lower execution time, as the replication case is only using half the processes for useful work. However, as we scale up, the checkpointing and rollback overheads become more prevalent. We can observe that the replication approach gives 13.04\% and 19.26\% reduction in execution times over checkpointing in CloverLeaf and PIC applications, respectively, at 8192 processes.

\begin{figure}
    \centering
    \includegraphics[width=\linewidth]{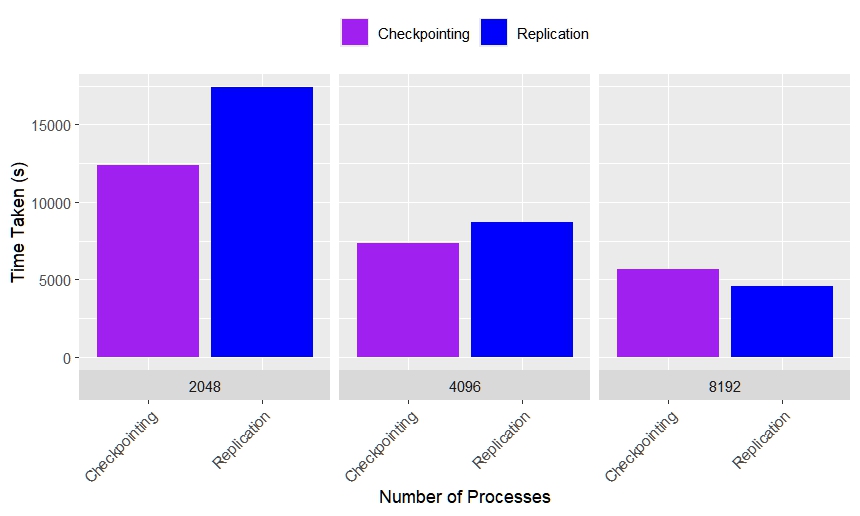}
    \caption{Comparison of Execution Times for Checkpointing and Replication for the PIC Simulation Skeleton Codes}
    \label{fig:pictime}
\end{figure}

\begin{figure}
    \centering
    \includegraphics[width=\linewidth]{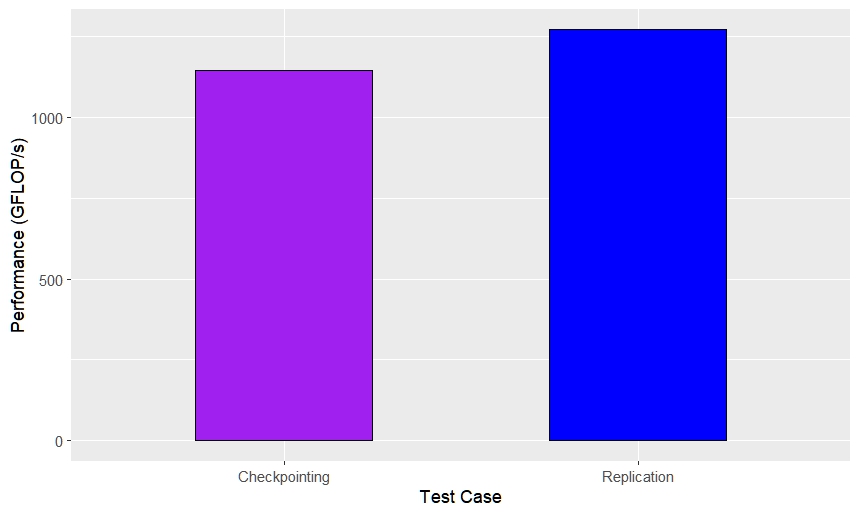}
    \caption{HPCG Performance under Log-Based Failure Simulations}
    \label{fig:logbased}
\end{figure}

Finally, to further validate the viability of replication-based fault tolerance, we have also obtained the HPCG performance under failures generated using the failure logs of the Tsubame 3 supercomputer \cite{taherin-examining-2021} with 8192 processes. These failure logs indicate the times when failures occurred in that system over the period from 2017 to 2020.
Tsubame 3 is a petascale system with a theoretical peak performance of 12.1 PFlop/s and an approximate MTBF of 230797 seconds. For our experiments, we scale the time between failures for each failure event by 100 to achieve an MTBF of 2308s, which is sufficient to have a noticeable impact on HPCG execution with a 3-hour target runtime. While we have reduced the overall timescale of the failures, we still maintain the relative timing of the failures as they occur in a real system. Unlike the probability-based failure simulations that considered process-level failures where the process chosen to be killed was randomly selected, the Tsubame 3 failure log contains node-level failures, and hence, with log-based failure simulations, the node chosen to be killed is determined by the log.
Each unique node name in the logs corresponds to a node in our 256-node execution. This accurately simulates scenarios in which the same node fails repeatedly. While node failures can have many causes, they actually manifest as a group of processes that become inaccessible to the rest of the job. All of our failure-handling mechanisms generally extend to a group of processes failing simultaneously and therefore can handle these node failures using the same general logic as for individual process failures. As we use the latter half of our set of processes as replicas, computational and replica processes generally exist on different nodes. When a node fails, if all the processes running on it have a corresponding replica running on a different node, execution can continue without any rollback. Figure \ref{fig:logbased} shows the HPCG performance using the log-based failure simulations. As in the previous probability-based results, the full-replication case performs better than checkpointing. The checkpoint/restart case performs slightly better with log-based failure simulations than with earlier probability-based ones, and is a bit more competitive with full replication due to its slightly higher MTBF of 2308 seconds compared to 2000 seconds and more spiky failures.

\subsection{Discussion}

Overall, our results validate the viability of replication as a fault-tolerance mechanism in large-scale systems through real executions. Furthermore, we observe that for certain execution times, full replication alone, i.e., without checkpointing, is sufficient to outperform pure checkpointing. We expect a similar pattern to hold at higher MTBFs, where pure checkpointing performs better at very low execution times, followed by full replication outperforming it as execution times become long enough to be substantially impacted by failures, and further followed by a combination of checkpointing and replication performing best at even higher execution times, where the number of failures is high enough to have a reasonable chance of killing both copies of a process. We also observed that the log-based failure simulations resulted in more failures occurring in bursts than the probability-based simulations. This slightly favors the pure checkpointing mechanism as it leads to lower rollback overheads. However, another contributor to performance loss when using log-based simulations is node downtime. If the checkpointing case uses all the resources it allocates to run computational processes, it must wait for all failed nodes to restart/recover after each failure. If combined with replication, however, it is possible to restart immediately with a lower replication degree if both copies of a process fail. Therefore, node downtime has a much lower impact on a combined checkpointing and replication framework as compared to a pure checkpointing framework.
\section{Conclusions and Future Work}
\label{con-fut}

In this work, we presented our FTHP-MPI fault tolerance framework that provides fault tolerance using replication for native MPI libraries that do not have support for fault tolerance, thereby providing the benefits of both fault tolerance and high performance. We conducted actual experiments on three applications, namely, HPCG, PIC, and CloverLeaf applications. Our experiments show that for applications with small execution times, of the order of a few hours, replication alone without the use of checkpointing is sufficient and provides 10.94-19.25\% higher performance than traditional checkpointing approaches. We showed that better efficiency is achieved beyond a certain number of processors by using additional processors for replication rather than using all processors for application execution with checkpointing. We also showed that the additional overhead of using our library in failure-free conditions is less than 1.5\%.

As future work, we plan to investigate combining adaptive partial replication and adaptive checkpointing along with strategies for scheduling in adaptive replication. This is enabled by our replication framework, which allows processes to be dynamically replicated during execution rather than only at the start. We also plan to further expand our library by adding support for other MPI features, such as MPI-IO and one-sided communications.

\bibliography{citations}

\end{document}